\documentclass[twocolumn]{paper}
\pdfoutput=1
\usepackage[utf8]{inputenc}
\usepackage[margin=2.5cm]{geometry}
\usepackage{listings}
\usepackage{cite}
\usepackage{amsmath,amssymb,amsfonts}
\usepackage{algorithmic}
\usepackage{graphicx}
\usepackage{textcomp}
\usepackage{authblk}

\begin{document}

\title{METICULOUS: An FPGA-based Main Memory Emulator for System Software Studies}
\author[1]{Takahiro Hirofuchi}
\author[1]{Takaaki Fukai}
\author[1]{Akram Ben Ahmed}
\author[1]{Ryousei Takano}
\author[2]{Kento Sato}
\affil[1]{Digital Architecture Research Center, National Institute of Advanced Industrial Science and Technology (AIST)}
\affil[2]{RIKEN Center for Computational Science (R-CCS)}

\maketitle

\begin{abstract}
Due to the scaling problem of the DRAM technology,
non-volatile memory devices, which are based on different principle of operation than DRAM, are now being intensively developed to expand the main memory of computers. Disaggregated memory is also drawing attention as an emerging technology to scale up the main memory.
Although system software studies need to discuss management mechanisms for the new main memory designs incorporating such emerging memory systems, there are no feasible memory emulation mechanisms that efficiently work for large-scale, privileged programs such as operating systems and hypervisors.
In this paper, we propose an FPGA-based main memory emulator
for system software studies on new main memory systems.
It can emulate the main memory incorporating multiple memory regions with
different performance characteristics.
For the address region of each memory device, it emulates the latencies, bandwidths and bit-flip error rates of read/write operations, respectively.
The emulator is implemented at the hardware module of an off-the-self FPGA System-on-Chip (SoC) board.
It is carefully designed to be fully compatible with the AXI4 protocol so as to support multi-core CPU and multiple memory regions.
Any privileged/unprivileged software programs running on its powerful 64-bit CPU cores
can access emulated main memory devices at a practical speed through the exactly same interface as normal DRAM main memory.
Through experiments, we confirmed that the emulator transparently worked for CPU cores and successfully changed the performance of a memory region according to given emulation parameters; for example, the latencies measured by CPU cores was exactly proportional to the latencies inserted by the emulator, involving the minimum overhead of approximately 240 ns.
As a preliminary use case, we confirmed that the emulator allows us to change the bandwidth limit and the inserted latency individually for unmodified software programs, making discussions on latency sensitivity much easier.
\end{abstract}


\section{Introduction}

The DRAM technology, being used for the main memory of computers for decades,
are now facing serious problems.
Because a memory cell of DRAM, a microscopic capacitor, needs refresh energy to keep its electric charge~\cite{micro2010dram},
the energy consumption of a memory subsystem accounts for a substantial part of
the total energy consumption of a computer (e.g., more than 40\% in a server computer~\cite{energyieee}).
This results in difficulty in equipping the memory subsystem with more DRAM modules.
Moreover, the leak energy problem in DRAM circuits becomes more acute due to finer manufacturing process.
Thus, DRAM is unlikely able to meet the ever-growing demand of memory capacity;
the increase of the capacity of a DRAM module does not follow the intense performance improvement of recent processors and accelerators (i.e., memory capacity wall problem).

Non-volatile memory devices~\cite{nvmeml,nvmnanoreview,mramieee}, which are based on different principles of
operation than DRAM, are now being intensively developed to expand the main memory of computers~\cite{itrs2013,isca09arch,isca09durable,isca09mainpcm,pageplacement,pdram,sttmrammain}.
Disaggregated memory\cite{memdis}, enabled by latest coherent interconnections (e.g., CXL),
is also drawing attention as an emerging technology to scale up the main
memory~\cite{isca09disagg,hpca12dis,infiniswap,ofc2019dis,legoos,nsdi16dis,firesim,socc17rm,pond}. It allows expanding the main memory by attaching remote tiered memory.
For example, Intel Optane Data Center Persistent Memory Module (DCPMM) is a
non-volatile main memory module built on the 3D XPoint technology, which is a stack
array of resistive memory elements. The capacity of a DCPMM is an order of
magnitude larger than that of a DRAM module.
In the same manner as a DRAM module,
DCPMMs are connected to the main memory bus of a computer via its DIMM interface.
In the configuration mode called AppDirect, the memory controller directly maps
the memory space of DCPMM to the physical address space of CPUs.
Software programs access its memory space by using the same CPU instructions as those of DRAM.

However, the performance characteristics of such emerging memory systems are
substantially different from those of today's normal main memory. For example, we observed that
the latency of read-only memory access of DCPMM was 374 ns and
the latency of memory access involving the writeback of cachelines was 391 ns,
which are 4 times and 4.1 times larger than those of DRAM, respectively~\cite{ieicedcpm}.
The bandwidth of sequential read access was 37 GB/s and
the bandwidth of writeback-involving access was 2.9 GB/s,
which were 37\% and only 8\% of those of DRAM, respectively.
Although measurement tools and tested systems were different, other work also
reported that DCPMM performs in a significantly different manner from DRAM \cite{dcpmmucsd1903,dcpmmmut1904}.
Other types of main memory devices such as MRAM and ReRAM are expected to be available in the market in the near future,
which will also have different performance characteristics.
Similarly, CXL-based remote memory is considered to incur add several hundreds nanoseconds due to the communication overhead of the interconnection.

The advent of such new types of main memory systems poses new challenges to system software studies.
Since the read/write performance of emerging memory devices is generally inferior to that of normal DRAM,
both DRAM and an emerging memory device are combined to expand the physical address space of the main memory.
This hybrid design of memory subsystems introduces a huge performance gap between the memory region of DRAM and that of the emerging memory device.
In addition, the memory region of an emerging memory device tends to involve prominent asymmetry in read/write performance.
These characteristics of the main memory have been never seen in DRAM-based memory systems.
Now, we are tackling open questions on how such new main memory systems should be designed and how they should be managed by system software programs such operating systems and hypervisors.
We need to clarify through quantitative evaluation,
how much performance improvement is possible by a proposed mechanism/algorithm.
We need to also discuss to what extent the proposed mechanism/algorithm is effective for what ranges of latency and bandwidth values of a target memory device;
in other words, in order to maximize performance of a particular software mechanism, what ranges of latency and bandwidth values need to be achieved by a target memory device.

However, there are no feasible emulation mechanisms of the main memory of
computers that efficiently work for system software programs.
We need an emulation mechanism supporting large-scale, privileged programs such
as operating systems and hypervisors.
Microarchitectural simulators, typically used for computer architecture studies,
are overwhelmingly slow to do a full system simulation.
Software-based emulation mechanisms using delay injection techniques (e.g.,
Quartz \cite{quartz} and our MESMERIC \cite{koshiba2019}) do not support privileged programs.
In the system software area, prior studies typically devised makeshift methods
to emulate larger latencies of memory devices,
which are far from how real memory devices behave.

In this paper, we propose an FPGA-based main memory emulator intended to be
used for system software studies on new main memory systems.
It can emulate the main memory composed of multiple memory devices.
For the address region of each memory device, it emulates the latency, bandwidth and bit-flip errors of read/write operations, respectively.
The emulation mechanism is implemented at the FPGA module of an FPGA System-on-Chip (SoC) board.
According to given emulation parameters,
it controls the queuing of read/write operations at the hardware layer.
It is carefully designed to be fully compatible with the AXI4 protocol so as to support multi-core CPU and multiple memory regions.
System software programs can access emulated main memory devices in the same manner as normal DRAM devices;
CPU cores access emulated devices by using load/store instructions via CPU caches.
Thanks to powerful 64-bit ARM CPU cores on the FPGA SoC board, full-fledged
operating systems run on the emulator, which enables system software
researchers to easily implement and evaluate new ideas for emerging memory
devices.
The emulator is designed for widely-available, off-the-shelf FPGA SoC boards;
the board of our prototype costs 1000 to 1500 USD, being affordable for the community.

Section 2 introduces the background of this study.
After discussing requirements in Section 3,
Section 4 presents the design of the proposed mechanism and its prototype implementation.
In Section 5, we evaluate the accuracy of the emulator and also report its use cases.
Section 6 summarizes related work. Section 7 concludes this paper.

\section{Background}
\label{sec:bg}

Recently, system software studies have been conducted for the new design of the
main memory that incorporates emerging memory devices.
For example, \cite{pmfs} and \cite{bpfs} proposed new file systems for byte-addressable, non-volatile memory devices.
The file system of an operating system performs read/write operations for an emerging memory device by using normal load/store instructions of CPU, not by the DMA mechanism of an I/O bus.
In \cite{pvm}, the memory region of a byte-addressable memory device is managed as a dedicated memory domain
in the same manner as remote memory in an NUMA architecture.
In \cite{nvheap}, a programming library safely manages memory objects allocated in such memory devices
in order to avoid inconsistency upon unexpected power outages.
\cite{mnemosyne} proposed a programming interface such as primitive data types to
control persistency for non-volatile main memory.
Yet, as system software researchers,
we proposed a hypervisor-based virtualization mechanism (RAMinate) for hybrid main memory systems \cite{socc2016raminate}.
It automatically optimizes page locations between memory devices; it detects
hot memory pages and relocates them to the faster memory device (i.e., DRAM in
general). RAMinate is implemented as a hypervisor and there is no need to modify guest operating systems.

The release of Intel Optane DCPMM and the expectation of succeeding similar
memory devices increase the importance of system software studies addressing
hybrid main memory systems. We need to emulate the performance of emerging main
memory systems for quantitative evaluation of proposed mechanisms.
However, as far as we know, there is no emulation mechanism that enables us to
efficiently execute large-scale, privileged programs.

\begin{figure}
	\includegraphics[width=\linewidth]{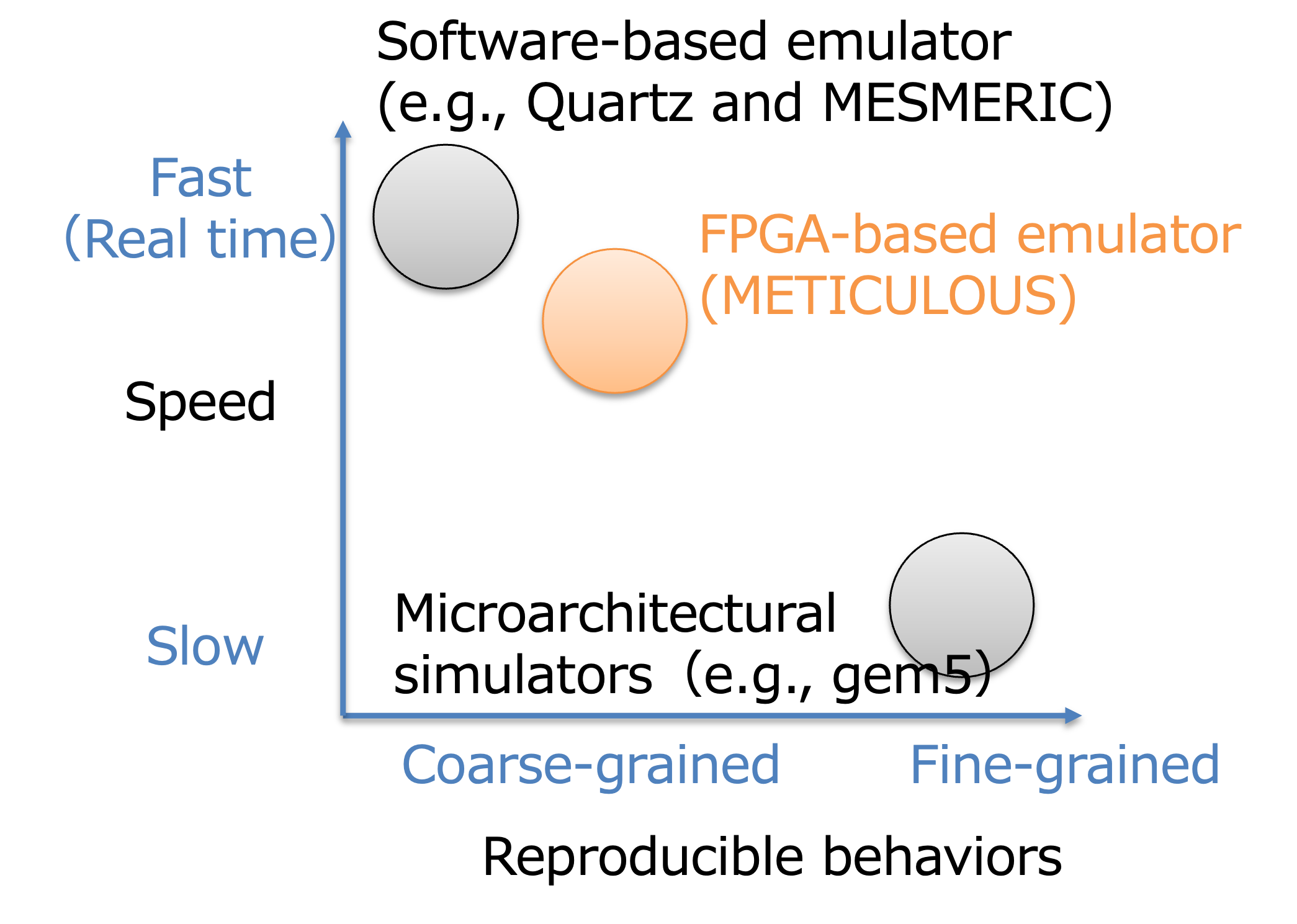}
	\caption{Conceptual trade-off in emulation approaches}
	\label{fig:tradeoff}
\end{figure}
Figure \ref{fig:tradeoff} illustrates a conceptual trade-off between emulation
speeds and reproducible behaviors.
In the computer science area, microarchitectural simulators~\cite{gem5,marss,zsim}) are
typically used for performance evaluation. It simulates the behaviors of internal states of processors.
For memory subsystem studies, they are often combined with memory device simulators~\cite{dramsim2,nvmain2,nvsim}. Although it can reproduce fine-grained behaviors of a hybrid main memory system,
the execution speed of a target program is enormously slow; in our past experiment, it took 8 hours to simulate only 1-second behaviors of a system. Thus, {\it kernel} workloads (i.e., a tiny program that models the characteristic behavior of a target workload) are typically used for the simulator.
Since operating systems and hypervisors are large-scale, general-purpose programs, it is difficult to define such {\it kernel} workloads. Although some microarchitectural simulators can execute an operating system, their execution speeds are not feasible for system software studies.

Software-based emulators (e.g., HPE's Quartz and our MESMERIC) can be used for performance
evaluation of userland workloads accessing an emerging memory device. They emulate
the large latency of such a memory device by using a normal DRAM-based computer.
They periodically retrieve performance monitoring events (e.g., cache miss information) from CPUs and dynamically adjust the speed of a target workload as if it is running on the slower memory device than DRAM. Our MESMERIC incorporates write-back activities of CPUs, so that it can more accurately emulates typical non-volatile memory devices having a huge performance gap between read and write operations.
Although these emulators do not suffer from simulation overheads, reproducible
behaviors of memory devices are coarse-grained; it does not reproduce the behavior of each memory request.
These mechanisms are not capable of supporting hybrid main memory systems, for which the target address of each memory request needs to be incorporated for emulation. Moreover, the software-based emulators, being implemented as userland mechanisms, are not applicable for evaluating the performance of new system software mechanisms implemented in an operating system kernel or a hypervisor.

Although being not illustrated in Figure \ref{fig:tradeoff}, binary instrumentation mechanisms (e.g., PIN\cite{pin}) are substantially slow due to the intercept of CPU instructions. It is also incapable of executing privileged programs.

\section{Requirements}
\label{sec:req}

We summarize requirements for the emulation mechanism of emerging main memory
devices, which is targeting for system software studies.

\begin{itemize}
	\item An emulator virtually creates the behavior of an emerging memory device
		incorporated into the main memory of a computer. It adjusts the
		performance parameters of the memory device such as read/write
		latencies, bandwidths and error rates as requested by users.
	\item An emulator executes large-scale, privileged, complex system
		software programs (e.g., operating systems and hypervisors) at
		acceptable execution speeds.
	\item An emulator can create hybrid main memory, which is now being common
		in hardware designs. It maps each memory device to a different
		region of memory address space and adjusts performance
		parameters for each memory region.
\end{itemize}

\section{Proposed Mechanism}

\begin{figure}
	\includegraphics[width=\linewidth]{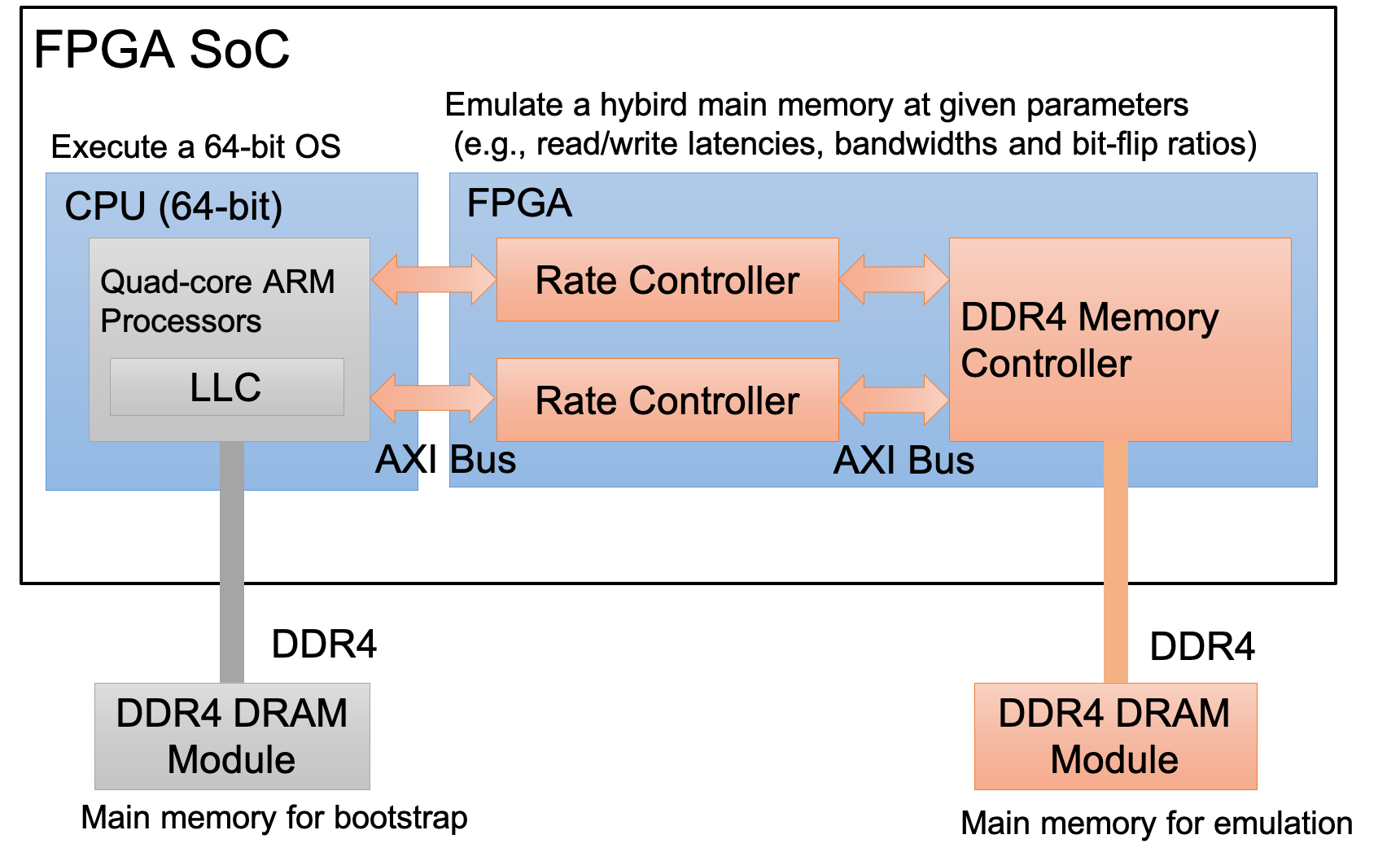}
	\caption{The overview of the FPGA-based Main Memory Emulator (METICULOUS)}
	\label{fig:overview}
\end{figure}

We propose an FPGA-based main memory emulator that enables us to evaluate the
performance of new system software mechanisms for emerging hybrid main memory
systems. As shown in Figure \ref{fig:tradeoff}, our emulator is intended to fill
the gap between microarchitectural CPU simulators and software-based emulators.
It can execute large-scale, privileged programs at acceptable speeds while
emulating the behavior of each memory request.
Figure \ref{fig:overview} illustrate the overview of the proposed mechanism, named METICULOUS. It uses a state-of-the-art FPGA System-on-Chip (SOC), which integrates a powerful 64-bits CPU and a programmable hardware module (i.e., FPGA). The CPU can execute full-fledged operating systems (e.g., Linux). The function of the FPGA is programmable by a hardware description language (HDL).
In the proposed mechanism, the FPGA is programmed to provide
an emulated hybrid main memory for the CPU. It controls the read/write latencies, bandwidths and
bit-flip error ratios to emulate the performance of memory devices.
It maps a DRAM module connected to the FPGA, called a FPGA-side DRAM module in this paper, into the physical address space of the main memory.
The CPU performs read/write operations to the FPGA-side DRAM module by issuing load/store instructions,
in the same manner as a normal DRAM module directly connected to the CPU (i.e., a CPU-side DRAM module).
The cache mechanisms of CPU (e.g., L1 and LLC) also works for the FPGA-side DRAM module.
From high-level design, the emulator has the rate controllers that adjust the performance of read/write operations to the FPGA-side DRAM module. Each rate controller is mapped to a part of the main memory.
It can insert a delay to the transfer of a read/write operation from/to is memory region, in order to emulate latencies and bandwidths. It can also manipulate data to emulate bit-flip
error ratios. We can dynamically change the performance parameters of each rate controller through its control registers. A program running on the CPU can access the registers that are mapped to a particular range of the physical address space.

As discussed in Section \ref{sec:eval}, the minimum latency of the FPGA-side
DRAM is approximately 400 ns due to the overhead of FPGA circuits; the emulator cannot
directly produce a latency value that is lower than it. However, this is not a problem for the emulator.
Since the emulator is designed to support multiple memory regions with different performance parameters,
we can reproduce the performance characteristics of a hybrid main memory system, by emulating relative performance differences between the memory regions.
We can set the performance parameters of one memory region to relative values against those of the other memory region.
In addition, it is possible to change the frequency of the CPU so as to adjust the
ratio of the processing capability of the CPU to the speed of its main memory.
In general, CPUs for server computers, performing aggressive out-of-order executions in many CPU cores, have relatively higher processing capability per memory bandwidth than those of embedded systems.

The memory mapping of the physical address space is flexibly configurable upon a system boot.
It is possible to map the memory regions of the emulator to any locations of the physical address space.
The boot loader reports to an operating system which physical address ranges are mapped to the emulator,
and then the operating system can incorporate them into its memory management system in any manner.
For example, if evaluating the latest NVDIMM support of an operating system for emerging memory devices,
we configure the operating system to incorporate the memory region of one emulated memory device (i.e., typically DRAM in mind) into its main memory and selectively use that of the other emulated memory device (i.e.., DCPMM or a future device) by its NVDIMM device driver (e.g., Device DAX in Linux).

The proposed mechanism supports the emulation of non-volatility of memory
devices. The emulator is capable of preserving written data to the main memory
upon system reboots, by keeping power supply to the FPGA-side memory.
The CPU instructions for cache management also work for the emulated main memory.

\section{Design}

\begin{figure*}
	\begin{center}
	\includegraphics[width=0.7\linewidth]{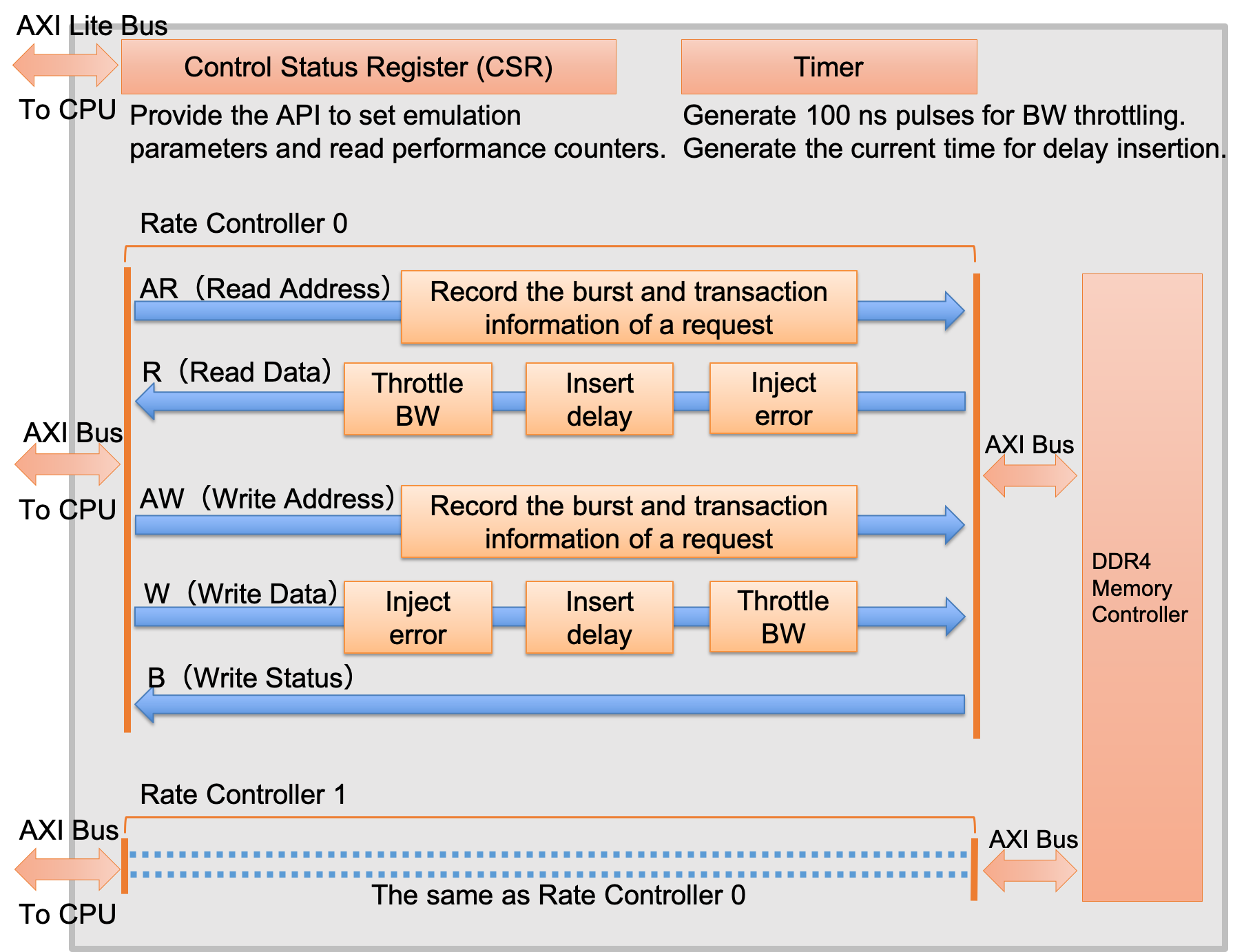}
	\end{center}
	\caption{Design Overview}
	\label{fig:design}
\end{figure*}

\subsection{Overview}

Figure \ref{fig:design} shows the design overview of METICULOUS, especially detailing its
FPGA side.
We carefully designed the emulator to have a high degree of modularity in it,
which contributes to enhancing future extensibility and also reducing development cost.
The CPU, the rate controller and the memory controller
are connected via a communication bus, respectively. It uses the AXI (Advanced
eXtensivle Interface) bus, which is a high-performance communication interface typically
used in ARM-based systems.
\begin{itemize}
	\item A rate controller is the key component in the emulator. It adjusts performance of read/write operations. It
		is implemented for each memory region of the emulator. Section \ref{sec:rate} explains its details.
	\item To ease development, the memory controller is implemented by using an existing IP
		module of a DDR4 DRAM controller, which is proprietary but
		available at no extra cost in many FPGA development frameworks.
		Because of the technical difficulty of controlling DRAM modules (e.g.,
		timing constraints on parallel signal lines), we avoid developing our own custom memory controller.
\item The control and status register (CSR) is to provide the CPU with the API to set
emulation parameters and read performance counters. Figure \ref{lst:api} details it.
We can dynamically set new performance parameters at any time. Even when read/write operations are ongoing, we can change latencies and bandwidths without any failure of operations. It is not necessary to restart the system when setting other performance parameters.
\item The timer is to generate 100-ns pulses distributed to the bandwidth throttling of each rate controller. It also generates the current clock distributed to the latency insertion of each rate controller.
\end{itemize}

\subsection{Rate Controller}
\label{sec:rate}

A rate controller is carefully designed to be fully compatible with the
protocol of the AXI4 bus. Because the CPU and the memory controller aggressively use the advanced features of the AXI bus to maximize performance,
we need to support all of them in the emulator; otherwise, the system hangs up.
\begin{itemize}
	\item Read/write requests and their replies are asynchronous; the master of an AXI bus (e.g., the CPU in the emulator) can issue the succeeding requests without waiting for the reply of the first request.
	\item Burst transfer needs to be supported; after the CPU sends the target address of a read/write request and its burst size,  multiple read/written data chunks are sequentially transferred as specified in the burst size.
	\item Reordering needs to be supported; each read/write request has a transaction ID, and requests having different transaction IDs can be reordered (i.e., in other words, the emulator needs to keep the order of the requests with the same transaction ID). We observed that each CPU core likely induces different transaction IDs. When the emulator has multiple memory regions, requests to a faster memory region is not allowed to outstrip those to a slower memory region if their transactions IDs are the same (i.e., one CPU core likely induced these requests).
\end{itemize}
We thoroughly verified the design of the rate controller. We believe that the rate controller can be transparently inserted into any AXI bus.

An AXI bus is composed of 5 communication channels and connected to its master and slave components.
For read operations, the rate controller works as follows:
\begin{itemize}
	\item The CPU first sends via the AR (Read Address) channel the
target address of a read request and other miscellaneous information such as a
burst size and a transaction ID. A rate controller looks up the CSR and finds
an emulation latency value of its memory region. It records the latency, its
burst size and a transaction ID in an ordered map structure. Its key is a transaction ID. For each transaction ID, it creates a FIFO queue to record a pair of a latency and a burst size).
\item Then, the memory controller returns a read data chunk via the R (Read) channel.
1) When error injection is necessary, the rate controller randomly flips the bits of read data according to the error ratio given in the CSR.
2) Then, the rate controller looks up the ordered map structure of the AR channel by the transaction ID of the read data chunk. According to its burst size, the rate controller enqueus the read data chunk and its succeeding read data chunks into a FIFO queue.
3) After a period of time for the latency has passed, the rate controller dequeues read data chunks.
When bandwidths throttling is necessary, the transmit of the read data chunks is done at the specified rate.
\item For bandwidth throttling, the token bucket algorithm is used to schedule transmissions.
For error injection, a liner feedback shift register is used to generate pseudo random numbers.
\end{itemize}
The rate controller similar works for the AW (Write Address) channel and the W (Write) channel. It passes through the B (Write Status) channel.

The latency insertion, bandwidth throttling and error injection mechanisms for the write channel are independent of those of the read channel. It allows setting emulation parameters for read and write operations, respectively.
For example, injecting errors only to the read channel is useful to emulate read disturbance in a non-volatile memory device. Sensing the resistance state of a cell may fail due to the lack of current supply. The memory controller incorrectly reads the value in memory cells, while the value in cells remains intact upon sensing.

\subsection{Memory Mapping}
The physical memory address offset that an FPGA-side DRAM is mapped to is defined as a
parameter in the FPGA circuit design. However, it is flexible how to split the
FPGA-side DRAM into multiple emulated memory devices.
A software program creates multiple memory regions in an FPGA-side DRAM, by
setting the start address and the size of each memory region in the CSR.
It is possible to dynamically change the offsets of memory regions without
restarting the FPGA board, as long as the operating system does not need
rebooting for new configurations.

It is the matter of the operating system how to treat those memory regions.
It should be tackled by system software studies.
For example,
in the case of Linux, we confirmed that we can assign its NVDIMM driver to them
and its persistent memory support (i.e, pmem) basically works.
We also succeeded in creating dedicated NUMA domains from them, without using its NVDIMM driver.
Further details are discussed in later sections.

\section{Implementation}
\label{sec:impl}

We developed the prototype of the emulator by using a middle-class Xilinx FPGA SoC board, ZCU104.
It has an FPGA SoC chip in which 4 ARM 64-bit CPU cores and an FPGA module are packaged.
It has 2 GB DRAM on board, directly connected to the CPU cores.
We added a 4 GB DRAM module to its SO-DIMM interface, which is connected to the FPGA module.
We also attached a 512 GB SSD module to its M.2 interface to set up software environments for experiments.
At the moment of writing, its total price in market costs only 1000 to 1500 USD.

For FPGA development, Xilinx Vivado v2019.2 was used.
The source code size of the rate controller is 952 lines in Verilog.
To test the design by simulation, 951 lines of SystemVerilog were also developed.

For software development,
Petalinux v2020.2 was used to create boot loaders (i.e., FSBR and u-boot), security
and power management firmware modules (i.e., BL31 and PMUFW) and Linux Kernel.
Since Linux's NVDIMM support for the 64-bit ARM architecture is not yet completed, we
intentionally chose the latest Petalinux to bump up the version of Linux
Kernel so as to test as many features as possible.

The theoretical bandwidth of the rate controller is 4.8 GB/s (i.e., the 128-bits bus runs at the frequency of 300 MHz). The used IP module of the DDR memory controller is connected through the bandwidth of 17.1 GB/s (i.e., 64 bits at 2133 Mbps). According to its product guide \cite{pg150}, the efficiency of the DDR bus is 23\% for random access. Even for the most severe access pattern (i.e., random access), our design is expected to achieve 3.9 GB/s (i.e., 23\% of 17.1 GB/s) as theoretical throughput.

\subsection{Configuration API}
We developed a thin software layer to configure the CSR of the emulator.
We tested the software library only on Linux but consider it portable to any other operating systems.
Listing \ref{lst:api} shows the excerpts of the configuration API.
The {\tt ME\_SetBoundary} function controls how to split the FPGA-side DRAM to multiple memory regions, by setting the start offset of each memory region.
The functions starting with the {\tt ME\_Set} prefix allow specifying latency, throughput and error rate for each memory region.
The API also provides functions to obtain statistical data such as the amount of transferred data and the number of bit errors.

We also developed a command line tool to wrap these functions. Users can easily
configure the parameters of the emulators from a console. They can also
automate experiments with different parameters by using it in a shell script.

\begin{figure*}
\lstset{language=C}
\begin{lstlisting}[
	caption={Excerpts of the configuration API functions of the emulator},
	captionpos=b,
	label=lst:api,
	frame=single,
	basicstyle=\tiny\ttfamily,
	]
/* Set the start offset of the memory region of an emulated device */
ME_SetBoundary(const unsigned int bank, const unsigned int boundary);

/* Set emulation parameters */
ME_SetLatency(const unsigned int bank, const unsigned int rd_latency_100ns, const unsigned int wr_latency_100ns);
ME_SetThroughput(const unsigned int bank, const unsigned int rd_thpt_10mbps, const unsigned int wr_thpt_10mbps);
ME_SetErrorRate(const unsigned int bank, const unsigned int rd_error_rate_in_percentage, const unsigned int wr_error_...

/* Get statistical data such as transferred bytes and injected bit-flips */
ME_GetXferredWrDataAmt(const unsigned int bank, unsigned long *data);
ME_GetXferredRdDataAmt(const unsigned int bank, unsigned long *data);
ME_GetRdBitErrors(const unsigned int bank, unsigned long *data);
ME_GetWrBitErrors(const unsigned int bank, unsigned long *data);
\end{lstlisting}
\label{fig:api}
\begin{flushleft}
The configuration API of the emulator, with prefix {\tt ME\_} standing for Memory Emulator, allows creating a memory region, setting emulation parameters and getting statistical data for each memory region. A memory region is called {\it a bank} in this API.
\end{flushleft}
\end{figure*}

\subsection{Device Tree}

We developed hardware descriptions for Device Tree to integrate memory regions into the memory management of Linux.
For architectures such as ARM, Linux obtains the hardware information of a computer through the pre-defined data structure called Device Tree.

\subsubsection{NVDIMM}
To use the memory regions of the emulator through the NVDIMM driver, the description of Listing \ref{lst:nvdimm} is enabled.
When the operating system boots, the {\tt pmem} driver of Linux is automatically attached to the physical memory regions of the emulator.
Users can create namespaces (i.e., logical memory regions at the operating system level) for each physical memory region, by the standard utility program of the NVDIMM driver ({\tt ndctl}).
We confirmed that the {\tt devdax} mode of the NVDIMM driver successfully works; users can create their device files (e.g., {\tt /dev/dax0} ) for namespaces.
Userland programs can map them into their virtual memory by the {\tt mmap()} system call and directly access them by bypassing all the I/O mechanisms of the operating system.
We also confirmed that the {\tt sector} mode works for namespaces on the emulator. These namespaces are used as block devices. Users can format a namespace with a file system such as ext4, mount it and perform any file operations.

We found that the tested Linux kernel (i.e., Linux 5.4 included in the used
PetaLinux) lacks the support of the {\tt fsdax} mode for the ARM architecture,
which allows users to create file systems on NVDIMM without the intervention of
the block device layer.
We expect that as ARM-based server computers become widely used in data
centers, the NVDIMM driver for the ARM architecture will be improved and the
limitation we have seen will be solved in future versions of Linux kernel.

The used FPGA SoC board does not support CPU instructions to control
persistency in a fine-grained manner, which have been introduced in the latest
ARM standard. Since the design of the emulator is portable, it will be possible to implement the emulator with another FPGA SoC board equipped with the latest architectural support for persistent memory.
Also, since the intention of the emulator is to evaluate the impact of memory device characteristics on system performance (not focusing on persistency control methods),
we consider that this limitation does not hinder the practicality of the emulator.

\begin{figure}
\lstset{language=C}
\begin{lstlisting}[
	caption={The description for Device Tree to use the memory regions of the emulator as NVDIMM},
	captionpos=b,
	label=lst:nvdimm,
	frame=single,
	basicstyle=\tiny\ttfamily,
	]
/* Make an emulated memory seen as NVDIMM */
amba: amba {
  #address-cells = <2>;
  #size-cells = <2>;
  pmem0: pmem@0x1000000000 {
    compatible = "pmem-region";
    /* offset 64GB, size 4GB */
    reg = <0x00000010 0x00000000 0x00000001 0x00000000>;
  };
};
\end{lstlisting}
\begin{flushleft}
A 4-GB NVDIMM region is created from the offset of 64 GB of the FPGA-side memory.
The {\tt pmem} driver is attached to it.
\end{flushleft}
\end{figure}

\subsubsection{NUMA}
The other option to integrate the memory regions of the emulator is the use of the NUMA support of Linux,
which is enabled by the description of Listing \ref{lst:numa}.
In this example, we created 3 NUMA domains. The first one is the CPU-side DRAM that is not treated by the emulator. All the CPU cores of the FPGA SoC board are located in this domain.
The second and third ones are the memory regions of the emulator, which will be assigned different performance characteristics.
As shown in Listing \ref{lst:numatopo}, the standard utility tool of Linux to control NUMA is used to observe the memory topology on the operating system.
Since emulated memory regions are transparently integrated, it is possible to enjoy rich features of the NUMA support of Linux.
All the CPU cores and the CPU-side DRAM are located in Domain 0, and the CPU cores are far from the NUMA Domains containing emulated memory regions; the Linux
kernel and userland programs do not use emulated NUMA domains unless explicitly specified to do so.
Users can exclusively assign their experimental program to an emulated NUMA domain by using {\tt numactl}, in order
to evaluate its performance sensitivity with difference memory device
characteristics.
In this example, we also demonstrate that the emulator can create multiple
memory regions. By changing parameters of each memory region individually,
users can investigate the feasibility of software mechanisms supporting hybrid
memory systems.

\begin{figure}
\lstset{language=C}
\begin{lstlisting}[
	caption={The description for Device Tree to use the memory regions of the emulator as NUMA domains},
	captionpos=b,
	label=lst:numa,
	frame=single,
	basicstyle=\tiny\ttfamily,
	]
/* NUMA Domain 1 and 2 are emulated */
// CPU-side DRAM (NUMA Domain 0)
memory@0x0 {
  device_type = "memory";
  reg = <0x0 0x0 0x0 0x7ff00000>;
  numa-node-id = <0>;
};

// FPGA-side DRAM (NUMA Domain 1)
memory@0x1000000000 {
  #address-cells = <2>;
  #size-cells = <2>;
  device_type = "memory";
  /* offset 64GB, size 2GB */
  reg = <0x00000010 0x00000000 0x00000000 0x80000000>;
  numa-node-id = <1>;
};

// FPGA-side DRAM (NUMA Domain 2)
memory@0x1080000000 {
  #address-cells = <2>;
  #size-cells = <2>;
  device_type = "memory";
  /* offset 66GB, size 2GB */
  reg = <0x00000010 0x80000000 0x00000000 0x80000000>;
  numa-node-id = <2>;
};

distance-map {
  compatible = "numa-distance-map-v1";
  distance-matrix = <0 0 10>,
                    <0 1 20>,
                    <0 2 20>,
                    <1 1 10>,
                    <1 2 20>,
                    <2 2 10>;
};
\end{lstlisting}
\begin{flushleft}
2 2-GB NUMA Domains are created from the offsets of 64 GB and 66 GB of the FPGA-side memory, respectively.
\end{flushleft}
\end{figure}

\begin{figure}
\lstset{language=C}
\begin{lstlisting}[
	caption={The operating system view of a NUMA topology},
	captionpos=b,
	label=lst:numatopo,
	frame=single,
	basicstyle=\tiny\ttfamily,
	]
\$ numactl -H
available: 3 nodes (0-2)
node 0 cpus: 0 1 2 3
node 0 size: 1933 MB
node 0 free: 1850 MB
node 1 cpus:
node 1 size: 2015 MB
node 1 free: 1984 MB
node 2 cpus:
node 2 size: 2003 MB
node 2 free: 1974 MB
node distances:
node   0   1   2
  0:  10  20  20
  1:  20  10  20
  2:  20  20  10
\end{lstlisting}
\begin{flushleft}
3 NUMA domains are observed from the operating system. Domain 0 has all the CPU cores and the CPU-side DRAM. Domain 1 and 2 contains memory regions on the emulator.
\end{flushleft}
\end{figure}

\section{Evaluation}
\label{sec:eval}

First, we evaluated the accuracy of the emulator through experiments using our microbenchmark programs.
Second, we applied the emulator to a study on application performance analysis in order to demonstrate its usefulness.
As for the requirements discussed in Section \ref{sec:req},
we confirmed that 1) the emulator successfully emulated latencies, bandwidths and error rates and 2) it transparently worked for the operating system in the same manner as the CPU-side DRAM does.
Although not presenting quantitative results, we also confirmed that 3) it succeeded in creating multiple memory regions and activated emulation parameters for each region. Thorough validation tests employing all the CPU cores and 2 memory regions were successfully passed.

\subsection{Microbenchmark}

To measure the bare hardware performance of the emulator without the interference of the operating system,
we did not use the NVDIMM and NUMA mechanisms mentioned in the previous section.
Instead, we developed programs to directly access the physical address range of the memory regions on the emulator.
These programs use the special device file ({\tt /dev/mem}) of Linux, which allows a userland
program to map any physical address range of a computer to the virtual
address space of the program. Since this mechanism is intended to be used for the register configuration of a memory-mapped device, the original Linux kernel disables CPU cache for the memory regions of the emulator. We modified the driver of {\tt /dev/mem} to allow a userland program to enable/disable CPU cache for the memory regions.
The used FPGA SoC board has 4 CPU cores of ARM Cortex-A53. Each CPU core has 32-KB L1 cache and 4 CPU cores share 1-MB L1 cache.

\subsubsection{Basic overhead}

We first discuss the bare latency of the emulator without inserting any delay in it.
We carefully designed a microbenchmark program to measure memory latency itself without the effect of CPU cache.
It first allocates a memory buffer and splits it into chunks of the cache line size (i.e., 64 bytes).
It then measures a period of time to access each chunk in the random order predefined in advance.
To minimize the effect of the prefetch and out-of-order execution of a CPU core,
the random order is implemented as a link list of chunks, so that a CPU core
cannot effectively predict the next address to be accessed by the program.
Every chunk access likely induces a cache line fetch from the main memory device.
The average time of this memory fetch is a memory latency.
When measuring a latency involving writeback (i.e., the eviction of a dirty cache line), the program writes an 8-byte value to each chunk upon its access. In this case, a CPU core performs memory fetch and writeback simultaneously.
For the details of the program, see our paper reporting the performance evaluation of Optane DCPM\cite{ieicedcpm}

\begin{figure}
\includegraphics[width=\linewidth]{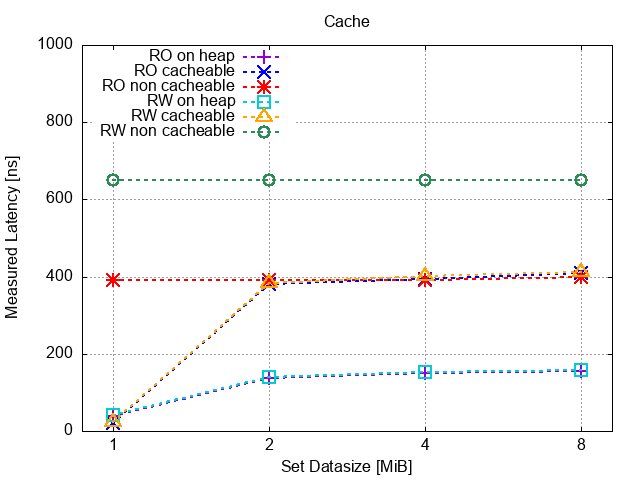}
\caption{Measured latencies with different buffer sizes}
\label{fig:cache}
\end{figure}

Figure \ref{fig:cache} shows measured latencies with different buffer sizes of the program.
For comparison, we also measured the latency of the CPU-side DRAM.
The read-only latency of the CPU-side DRAM ({\tt RO on heap}) and that
writeback-involving latency ({\tt RW on heap}) were approximately 160 ns,
when the buffer size was larger than the size of L2 cache (1 MB).
Those of the FPGA-side DRAM ({\tt RO on cacheable} and {\tt RW on cacheable}) were approximately 400 ns.
The overhead of the emulator, caused by AXI bus communications and the rate controller, caused the difference of 240 ns.

In this experiment, we also confirmed that
the FPGA-side DRAM worked with CPU cache in the same manner as the CPU-side DRAM;
when the buffer size was set to 1 MB (i.e., the same size as L1 cache),
both the above latencies were approximately 20-30 ns, thanks to CPU cache hits.
In contrast, in the case of the FPGA-side DRAM with CPU cache disabled ({\tt RO non cacheable} and {\tt RW non cacheable}),
the latencies did not change even when the buffer size was small.
Without CPU cache, the writeback involving latency of the FPGA-side DRAM increased by 240 ns,
which corresponds to the overhead of the flush of a dirty cache line.
Every store CPU instruction involved an actual memory write operation through the emulator.

\subsubsection{Latency control}
We evaluated the accuracy of the latency control mechanism of the emulator.
We configured the read/write delay parameters of the emulator to various values
and observed the actual latency values by using the above microbenchmark program.
The buffer size of the program was set to 4 MB to induce CPU cache miss.

\begin{figure}
\includegraphics[width=\linewidth]{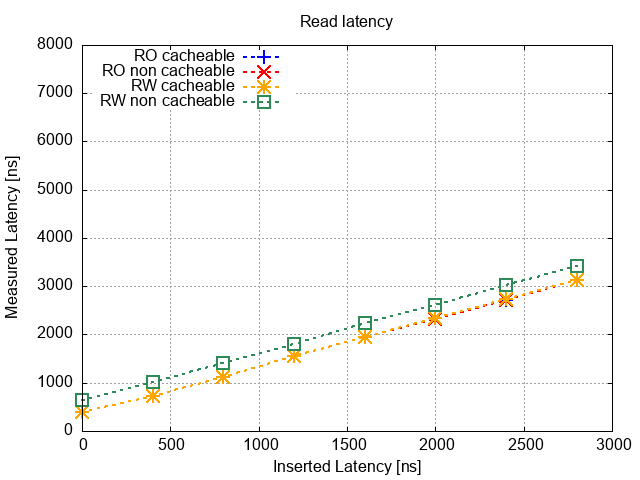}
\caption{Measured latencies with read delays inserted}
\label{fig:readlat}
\end{figure}

Figure \ref{fig:readlat} shows the results of the experiment in which only the
read delay was inserted by the emulator.
The measured latency proportionally increased by an added read delay.
For example, when 2000 ns was inserted by the emulator, the observed latency of a read-only operation ({\tt RO cacheable}) was
approximately 2400 ns, which was the inserted delay plus the bare
latency of the emulator (i.e., approximately 400 ns as already discussed).
The observed latency of involving writeback {\tt RW cacheable} was also approximately 2400 ns,
because the inserted read delay did not affect the performance of write operations.

\begin{figure}
\includegraphics[width=\linewidth]{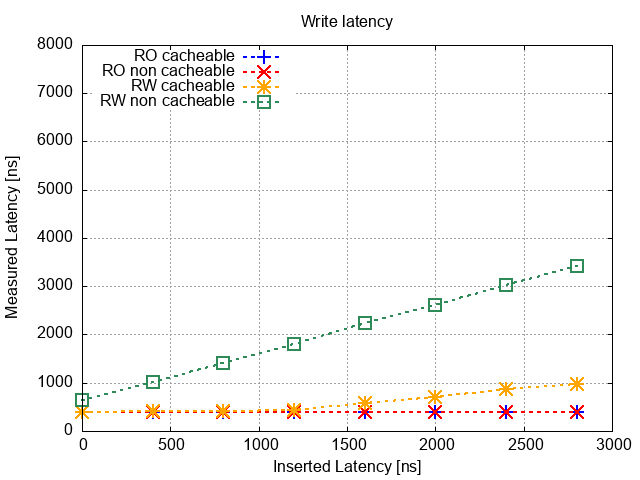}
\caption{Measured latencies with write delays inserted}
\label{fig:writelat}
\end{figure}

In Figure \ref{fig:writelat}, only the write delay was inserted and the read delay was not.
Clearly, the latency of a read only operation ({\tt RO cacheable}) was not affected by the inserted write delay.
The latency involving writeback ({\tt RW cacheable}) was not impacted when the
inserted write delay was small (i.e., 400 ns, 800 ns and 1200 ns),
and slowly increased as the delay was set to larger values.
The cache controller of CPU evicts modified cache lines asynchronously, which does not stop CPU cores executing instructions as long as the eviction queue is not full.
This result suggests that even though typical non-volatile memory devices suffer from large
write latency, performance degradation will be greatly alleviated for particular types of workloads taking advantages of asynchronous eviction.

\begin{figure}
\includegraphics[width=\linewidth]{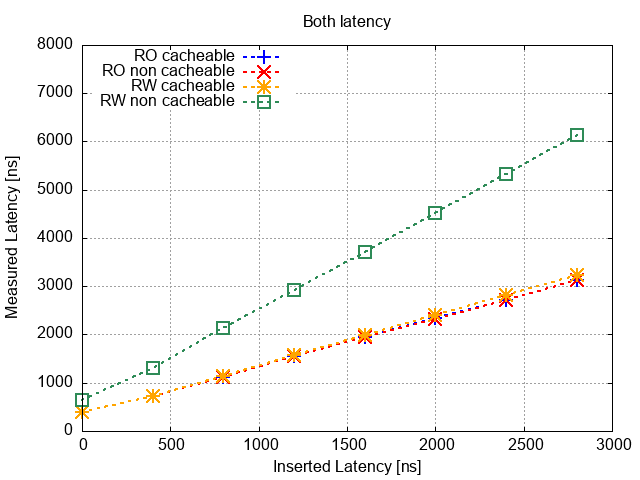}
\caption{Measured latencies with both read and write delays inserted}
\label{fig:bothlat}
\end{figure}

When both read and write delays were set at the same time, the control mechanism also correctly worked as in Figure \ref{fig:bothlat}.
The result captured both the features shown in Figures \ref{fig:readlat} and \ref{fig:writelat}.

In summary, we confirmed that the emulator correctly controlled read/write memory latencies, independently.
It also correctly worked in conjunction with the CPU cache mechanism of the existing CPU cores.
Due to the overhead of the emulator, 400 ns is the minimum latency, which is larger than that of the CPU-side DRAM.
It should be noted that application performance on the emulator is not intended to be compared with that on the CPU-side DRAM.
Users can evaluate the impact of memory latency by relatively comparing application performance with different delay parameters on the emulator. For studies on hybrid memory, the emulator supports also multiple memory regions to which different delay parameters are set.

\subsubsection{Bandwidth control}

To evaluate the accuracy of the bandwidth control, we developed a
microbenchmark program to measure the throughput of read and write data transfers at the emulator.
The program creates multiple threads performing sequential memory access.
In experiments,
because the used FPGA SoC board has 4 CPU cores, it creates 4 threads to maximize throughput.
Each thread allocates its own 1-MB buffer and performs read/write operations.
The program obtains through the configuration API of the emulator the amounts of transferred bytes of read and write data channels, and then calculates transferred bytes per second.

\begin{figure}
\includegraphics[width=\linewidth]{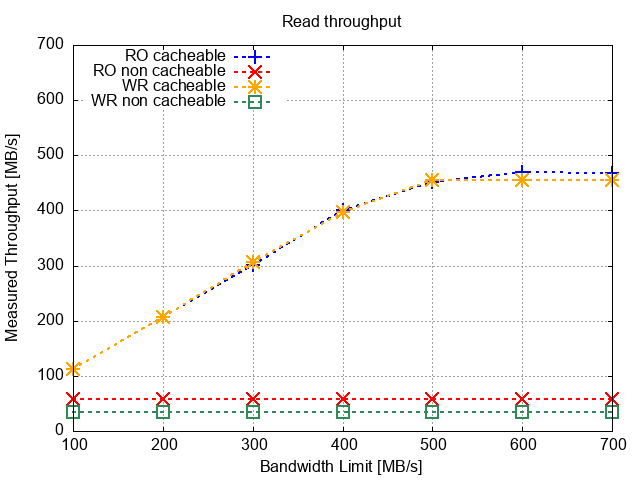}
\caption{Measured read throughput with read bandwidth limits}
\label{fig:readthpt}
\end{figure}

We first limited the read bandwidth of the emulator.
As shown in Figure \ref{fig:readthpt}, for both read-only and
writeback-involving workloads ({\tt RO cacheable} and {\tt RW cacheable}), the throughput of transferred read data through
the emulator was accurately limited to the specified read bandwidth values (i.e., 100, 200, 300 and 400 MB/s).
However, with higher bandwidth values such as 500 MB/s, the throughput did not increase beyond 450-460 MB/s, which is less than the theoretical one of the designs (discussed in Section \ref{sec:impl}).
We consider that this is the maximum throughput that the used FPGA SoC board can achieve for the emulator.
The number of memory requests that the CPU cores of the board can issue without waiting for their replies supposedly impacts on the maximum throughput. According to the reference manual of ARM Cortex A53 \cite{ddi0500d}, the issuing capability of the used FPGA SoC board is estimated to be 37 and 17 for read and write requests, respectively.

\begin{figure}
\includegraphics[width=\linewidth]{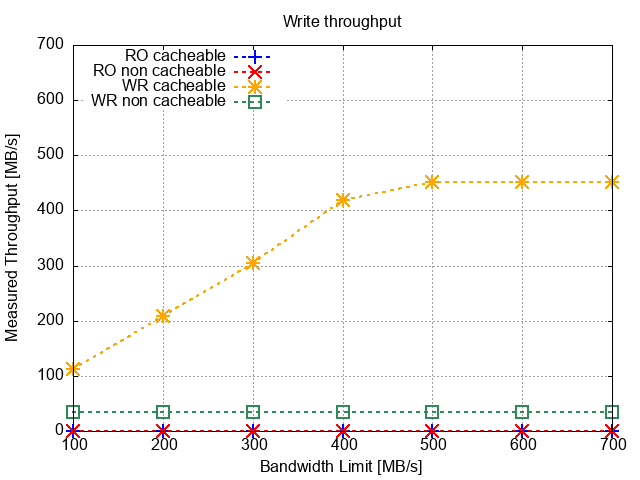}
\caption{Measured write throughput with write bandwidth limits}
\label{fig:writethpt}
\end{figure}

Similarly, in the case with the write bandwidth limited, the write throughput at the emulator
({\tt RW cacheable}) was accurately limited to the specified bandwidths up to 400
MB/s, as in Figure \ref{fig:writethpt}. Because read-only workloads did not involve write data transfers,
the lines of {\tt RO cacheable} and {\tt RO non cacheable} overlaps at zero.

In the above experiments, when CPU cache was disabled, read/write throughputs
were 50 and 60 MB/s for read-only and writeback-involving workloads ({\tt RO
non cacheable} and {\tt RW non cacheable}), respectively, which were under the
smallest specified bandwidth (i.e., 100 MB/s). If necessary, the emulator can limit
read/write bandwidths in a more fine-grained manner such as 10 MB/s intervals.

In summary, we confirmed that the emulator correctly limited the bandwidth of
read/write data transfers from/to the FPGA-side DRAM. With the FPGA SoC board
we tested, users can set any bandwidth capping values up to 450 MB/s.

\subsubsection{Error injection control}

We developed a program to verify the error injection of the emulator.
It first allocates a 4-MB buffer from the FPGA-side memory, which is initially
filled with zero on the emulator.
In the read-only mode of the program, it sequentially reads the buffer, counts the number of on bits and calculates its observed error rate. Similarly, in the write-read mode, it sequentially writes the buffer with zero and then reads the buffer again.
Although the emulator enables us to set any error rate of an order of $1/2^{32}$, in the below experiments, we intentionally set impractically high error rates (i.e., the order of 10\%) to ease verification.

\begin{figure}
\includegraphics[width=\linewidth]{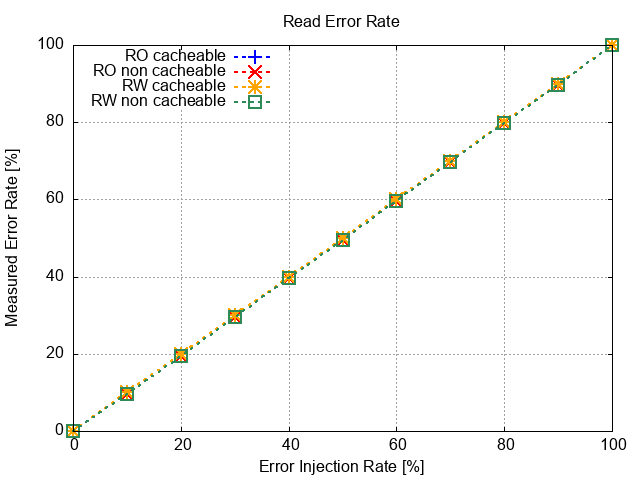}
\caption{Measured bit-flip errors with read error injections}
\label{fig:readerr}
\end{figure}

\begin{figure}
\includegraphics[width=\linewidth]{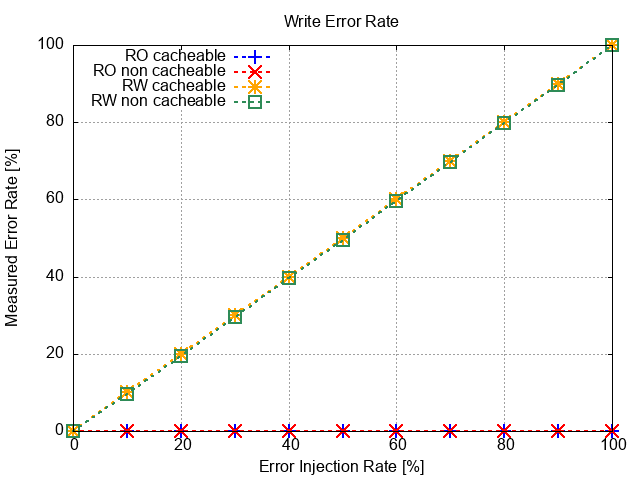}
\caption{Measured bit-flip errors with write error injections}
\label{fig:writeerr}
\end{figure}

Figure \ref{fig:readerr} shows the measured error rate with read error
injections. In either the read-only and write-read modes, the measured error
rates (e.g., {\tt RO cacheable} and {\tt RW cacheable}) correctly matched the tested read error rates set to the emulator.
As in Figure \ref{fig:writeerr}, in the case of write error injections, while not detecting any bit flips in the read-only mode, the program in the write-read mode (e.g., {\tt RW cacheable}) detected expected error rates for the set write error rates.
In the experiments, CPU cache did not contribute to mitigating observed bit-flip errors.
The buffer size for sequential access exceeded the CPU cache size.
If the memory access pattern of a workload efficiently takes the advantage of CPU cache, observed error rates with CPU cache will be smaller than those without it.

In summary, we confirmed that the emulator successfully performs error
injections to the read and write channels.
It enables users to set error rates for read and write separately,
which will contribute to emulating various error situations in memory devices such as read disturbance and program disturbance.

\subsection{Use case}

\begin{figure}
\centering
\includegraphics[width=0.7\linewidth]{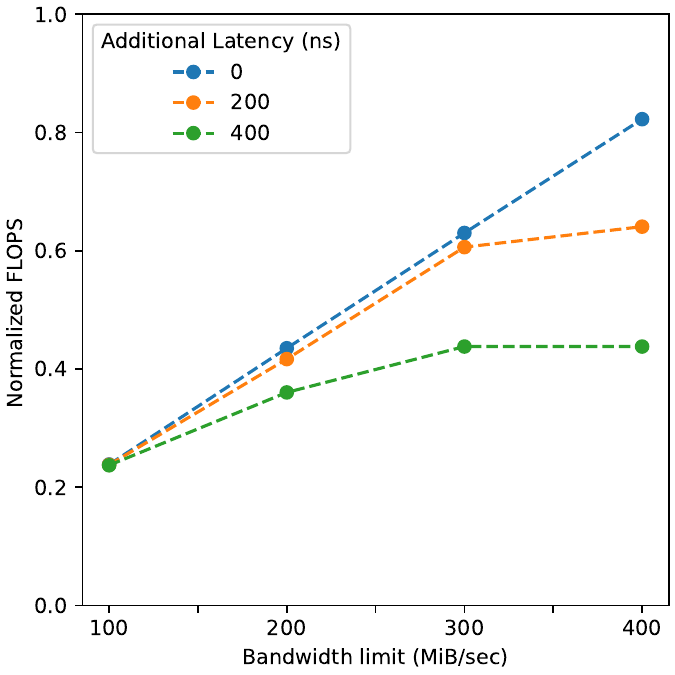}
\caption{Normalized FLOPS with emulated memory bandwidths and latencies (BabelStream)}
\label{fig:babelstream}
\end{figure}

\begin{figure}
\centering
\includegraphics[width=0.7\linewidth]{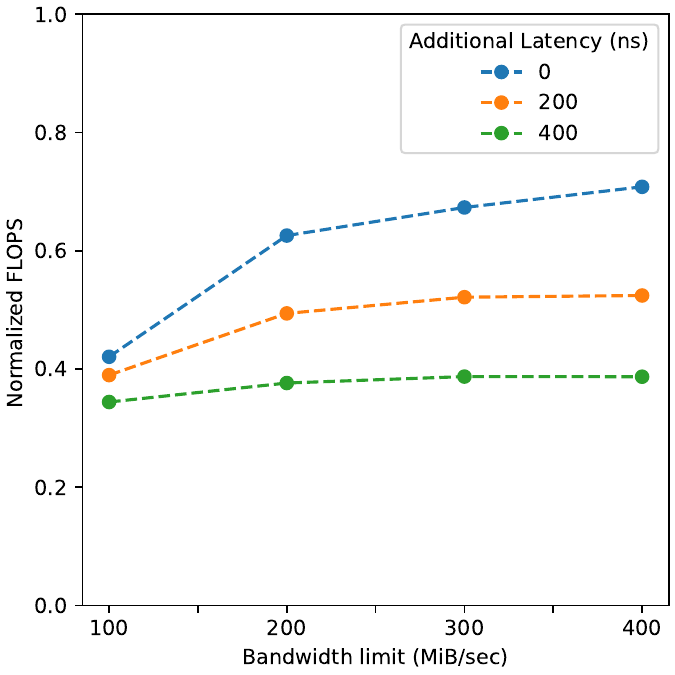}
\caption{Normalized FLOPS with emulated memory bandwidths and latencies (XSBench)}
\label{fig:xsbench}
\end{figure}

We started using the emulator to explore designs tradeoffs in the memory subsystem.
Historically, the roofline model \cite{Roofline} has been used to analyze theoretical and actual performance of computer systems.
In that, the memory access intensity of each application is modeled as a B/F (bytes per flop) number, i.e., the amount of data transferred from/to memory when executing one floating-point operation.
The roofline model can clearly illustrate whether performance bottleneck is located at the computation speed of a computer, at its memory bandwidth, or at another unknown component. Even in the case that memory latency is primary bottleneck, the roofline model cannot pinpoint it.
The advent of non-volatile memory devices and disaggregated memory systems increase the importance of discussing memory latency.
Our memory emulator,  allowing users to set memory latency and bandwidth independently, makes it possible to discuss how latency and bandwidth have impact on system performance. It is useful to explore the design space of a memory subsystem covering both the system software and hardware layers.

Having that in mind, we used the emulator to observe the latency sensitivity of programs running on the memory management mechanism of Linux-5.4.
We configured the emulator to set up one emulated memory region and attach it to the main memory as a NUMA node.
For target programs, we used the default memory management policy of the NUMA support of Linux,
which was specified through the {\tt numactl} command. Thanks to the transparency of the emulator, no modification to target programs was necessary.

Figures \ref{fig:babelstream} and \ref{fig:xsbench} show the results of latency sensitivity of tested target programs, BabelStream and XSBench.
We changed memory bandwidth limits in different inserted latencies (i.e., 0, 200 and 400ns).
The measured value of floating operations per second (FLOPS) of each target program was normalized by the one with no bandwidth limit and latency insertion.
As shown in Figure \ref{fig:babelstream}, in the case of no additional latency, the performance of BabelStream, which is known as memory bandwidth intensive due to its massive dot product calculation\cite{domke}, proportionally related to bandwidths limits; the memory bandwidth was performance bottleneck. Even when memory latency was increased by 200 ns, the program was not sensitive to the increased latency as long as the memory bandwidth limit was 300 MB/s or less. However, the program suffered from the latency when the bandwidth limit was 400 MB/s and more.
In Figure \ref{fig:xsbench}, the result of XSBench, simulating neuron transport by the Monte Carlo method, revealed different latency sensitivity; the performance of the program was greatly impacted by the inserted latencies, not by the bandwidth limits, possibly due to its random memory access pattern.
Although this is a preliminary experiment, we confirmed that the emulator allows us to change the bandwidth limit and the inserted latency individually for unmodified software programs, thereby making discussions on latency sensitivity much easier. Such results will contribute to finding an optimal balance between the memory bandwidth and latency in the hardware design of a memory subsystem and also suggesting implications to design an algorithm of memory management performed at the system software layer.

\section{Related Work}

As discussed in Section \ref{sec:bg}, microarchitectural simulators and
software-based emulators are not sufficient for system software studies on
emerging memory device, which need to evaluate large, privileged programs.
Although hardware-based emulation has potential to fill the gap among such
existing techniques, few hardware-based emulators were developed in prior studies.

\cite{pmfs,msstucsd} used the specially customized hardware platform developed by Intel to emulate the performance of the main memory using a non-volatile memory device. They customized the processor microcode and firmware of an Intel Xeon-based server. It can insert additional CPU stalls upon a cache miss to emulate a read latency and can also limit the throughput of the memory controller to emulate read/write bandwidths.
However, this mechanism is not publicly available. Only a few researchers had a chance to use it for their studies. Although its detailed information is not disclosed, apparently the emulation of write latencies and error injections are not supported. Theoretically, it is possible to implement these features in this platform. Having said that, it is hard to imagine that Intel discloses their proprietary technologies to allow any researchers to extend it for their own purposes.

In contrast, the design of our emulator is open and portable to any other FPGA platforms.
The built image for the tested FPGA SoC board is available for the community.
For use cases where CPUs other than the ARM Cortex A53 processor of the tested board need to be used,
we ported the design to Xilinx Alveo U250, which is a more powerful FPGA board
with a PCI Express interface. We attached the Alveo board to an ordinary server machine with Intel Xeon processors and confirmed that its main memory was extended with the 64-GB FPGA-side DRAM capable of performance emulation. Furthermore, for studies investigating hardware and software co-designs for emerging memory devices, it is possible to implement a customized processor at the FPGA side by using for example an open source RISC-V processor.

\cite{omori} developed a main memory emulator using an FPGA SoC
board. In a similar way as our emulator, they use the onboard DRAM of the FPGA
SoC board for emulation. In their mechanism, in order to increase the latency of memory access,
the control timings of the onboard DRAM (e.g., tRCD - a period of time from the open of a row to a column read/write) are adjusted by modifying the RTL code of the memory controller.
We consider that even if the license of the memory controller IP allowed modifying its RTL code, it would be unsupported by the vendor to change these timings.
\cite{tuna} also developed a latency emulation mechanism on an FPGA SoC board. It briefly states that they delayed handshake signals of the AXI bus of the FPGA-side DRAM.
Although both the papers did not describe much detail of their design, apparently their mechanisms
seem to support only very basic latency emulations.
These studies used the Xilinx Zynq-7000 SoC platform. It is an entry-level product for cost sensitive markets. It is equipped with only dual 32-bit CPU cores with limited capabilities.
In contrary, our emulator is designed for more powerful platforms (e.g., Zynq UltraScale+).
It emulates not only latency but also bandwidth and bit-flip errors for multiple memory regions and supports the advanced transfers of the AXI4 protocol issued by full-fledged 64-bit CPU cores.

\section{Conclusion}
We proposed an FPGA-based main memory emulator
for system software studies, which can emulate the main memory incorporating multiple memory devices with
different performance characteristics on latency, bandwidth and error rate.
Through experiments, we confirmed that the emulator transparently worked for CPU cores and successfully changed the performance of a memory region according to given emulation parameters; for example, the latencies measured by CPU cores was exactly proportional to the latencies inserted by the emulator, involving the minimum overhead of approximately 240 ns.
As a preliminary use case, we confirmed that the emulator allows us to change the bandwidth limit and the inserted latency individually for unmodified software programs, making discussions on latency sensitivity much easier.

This work was partially supported by JSPS KAKENHI 19H01108.

\bibliographystyle{unsrt}
\bibliography{grivon}

\end{document}